\documentclass{elsart}

\usepackage{amssymb}
\usepackage{epsf,colordvi,color,amsbsy}
\usepackage[dvips]{graphicx}
\begin{document}

\begin{frontmatter}

\title{On the Solar Component in the Observed Global Temperature Anomalies}

\author[]{Stefano Sello \corauthref{}}

\corauth[]{stefano.sello@enel.it}

\address{Mathematical and Physical Models, Enel Research, Pisa - Italy}

\begin{abstract}

In this paper, starting from the updated time series of global temperature anomalies, Ta, we show how the solar component affects the observed behavior using, as an indicator of solar activity, the Solar Sunspot Number SSN. The results that are found clearly show that the solar component has an important role and affects significantly the current observed stationary behavior of global temperature anomalies. The solar activity behavior and its future role will therefore be decisive in determining whether or not the restart of the increase of temperature anomalies observed since 1975 will occur.

\end{abstract}
\end{frontmatter}

\section{Introduction}

It is well known that the Sun is the main input energy source of the Earth, thus the question of whether, and to what extent, the Earth's global climate is influenced by solar variability remains central to the current understanding of anthropogenic climate change. Some studies consider that the Sun total irradiance has long been stable and the current global warming is mainly attributed to the enhanced greenhouse gas effect due to anthropogenic emissions such as $CO_2$, $CH_4$ and $NO_x$. However, other studies consider that long-term solar variations may have been responsible for at least 30-50\% of the warming observed since 1975 and about 80\% since the Maunder solar minimum occurred in the mid 17th century (Scafetta and West, 2006; Scafetta, 2011). Other recent works consider the spectral solar irradiance to adequately consider the variations of the solar activity and their role in the behavior of global temperature anomalies (Feng-Ling Gao et al., 2015). We refer the reader to a good review on the subject, given by Lockwood, 2012, on the effects of solar activity on global and regional climates. \\
In the present study we address the problem of determining, on a pure statistical basis using time-series analyses, the role of the solar activity contribution on the observed behavior of global temperature anomalies, considering the current updated records.

\section{Time series of temperature anomalies and Sunspot Numbers}

The climatological time series used in this study is the monthly average of global temperature anomalies, Ta, provided by the Climatic Research Unit (University of East Anglia) in conjunction with the Hadley Centre (UK Met Office). The file analyzed is CRUTEM4 in which we calculate the average global hemispheric temperature land area as: $(2NH + SH) / 3$, weighted in order to prevent that the value becoming dominated by the Northern hemisphere, where there is typically greater observational coverage.\\
The file is updated periodically and can be dowloaded from the site: \\
\  http://www.metoffice.gov.uk/hadobs/crutem4/data/diagnostics/global/nh+sh/index.html\

The time series for monitoring the solar activity is the monthly averaged  Solar Sunspot Number, SSN provided by the WDC-SILSO, Royal Observatory of Belgium, Brussels. The choice of this parameter is mainly due to its wide temporal coverage, necessary for statistical studies. The file, regularly updated, and is available from:\  http://sidc.oma.be/silso/datafiles.\
The two above analyzed time series cover the time interval: January 1856 - May 2015 (1913 values).

 In figure 1 it is shown the first time series: CRUTEM4.

\begin{figure}[!ht]
\centering
\includegraphics[scale=1]{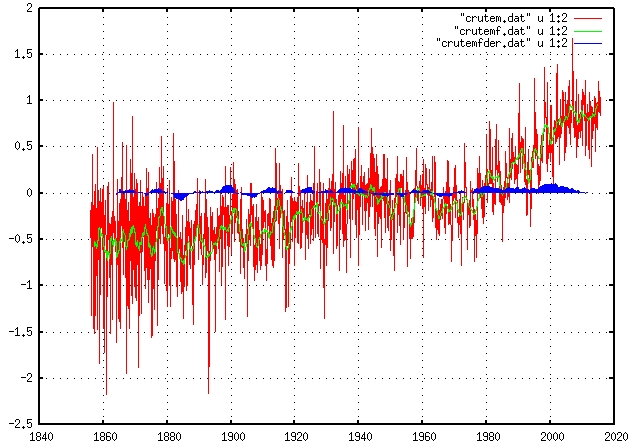}
 \caption{Red line: Global temperature anomalies, CRUTEM4 averaged over a monthly basis provided by the Climatic Research Unit (University of East Anglia). Green line: Temperature anomalies filtered with a moving average window of 24 months. Blue areas: First derivative, properly scaled, of the filtered temperature anomalies.}
 \label{fig1}
\end{figure}

In order to point out the long term trend of the time series, we performed a proper moving window low-pass filter with one step with amplitude 24 months as shown by the green line in Figure 1.
The blue areas in the same figure show the first derivatives, properly scaled, to highlight the periods of the different behavior of the series. There are three main periods: from 1856 to 1910, with a globally stationary trend; 1910-1940 with an increasing behavior; from 1940 to 1975 with an almost stationary behavior, from 1975 to 2007 with a strong temperature growth and from 2007 to present with a stationary behavior.

To better point out these various periods and to better identify any possible characteristic periodic or quasi-periodic phenomena we performed  a proper wavelet analysis of the filtered series. Figure 2 shows the result.

\begin{figure}[!ht]
\centering
\includegraphics[scale=1]{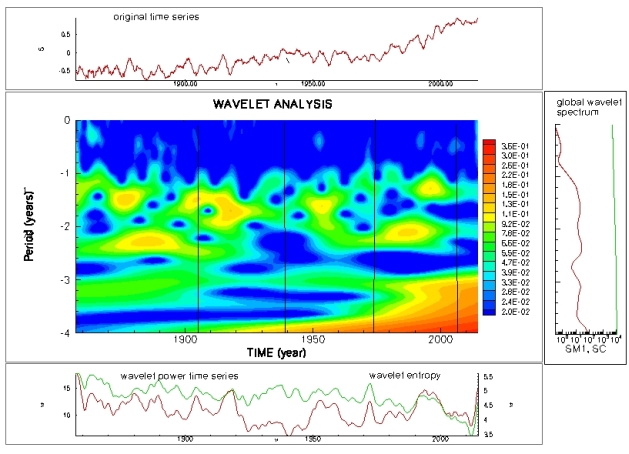}
 \caption{Wavelet analysis of global temperature anomalies, CRUTEM4, filtered with a moving window of 24 months. Note that the frequency scale is logarithmic.}
 \label{fig2}
\end{figure}

As we can see, some characteristic frequencies or local periodicities identified at various phases, such as 8-9 years and 4-5 years, can be evidenced and monitored to determine whether they have some predictive character.

In figure 3 it is shown the second analized time series: SSN.

\begin{figure}[!h]
\centering
\includegraphics[scale=1]{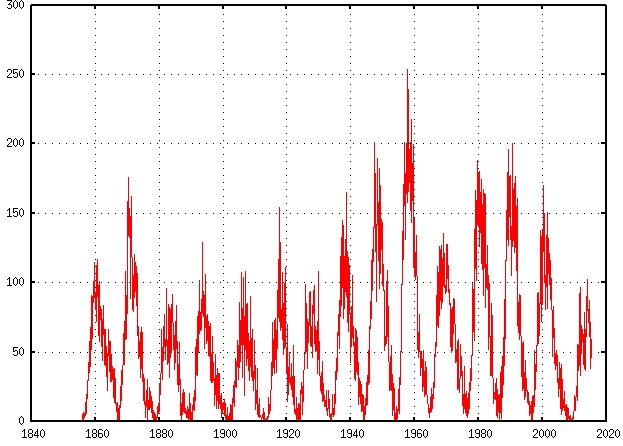}
 \caption{Solar Sunspot Number, SSN,  monthly averaged from WDC-SILSO, Royal Observatory of Belgium, Brussels.}
 \label{fig3}
\end{figure}

\section{Correlation analysis}

In order to perform a correlation analysis it is appropriate, starting from the original data, to apply a moving average window filter with a width of 96 months for the Ta series and a moving average window filter with a width of 132 months for the SSN series, in order to point out only the long-term overall behavior, excluding the high-frequency fluctuations which are not useful in determing the correlations we are looking for.

In figure 4 it is shown the trend of the two filtered series.

\begin{figure}[!h]
\centering
\includegraphics[scale=1]{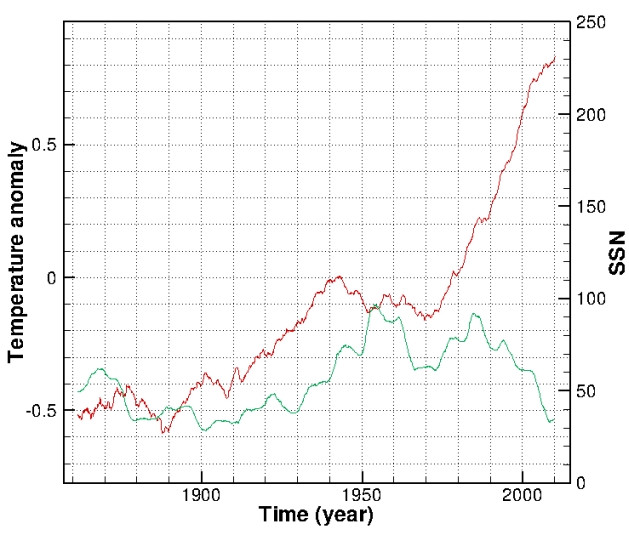}
 \caption{Red line: Ta series, filtered through a moving window of 96 months. Green line: SSN series filtered through a moving window of 132 months.}
 \label{fig4}
\end{figure}

The correlation analysis of the two series shows that the maximum correlation is when the series Ta is time shifted about 9 years (delayed) with respect to the series SSN, with a value of linear correlation coefficient of 0.60.
If we observe in Figure 5 the trend of the nine years shifted filtered series, we can note that the two series: SSN and Ta, show a trend quite coherent, in phase until around 1986. After this date, the trend of SSN is reversed with respect to Ta, with gradually decreasing values, while the Ta continues to grow with the same rate. The loss of coherence is due to the extra-solar components that determine the resulting trend in this phase. This fact introduces an additional shift or time delay in the response of Ta to SSN behavior. In fact, a change in the Ta begins to occur only around 2000 where we can see an inflection point in the related curve, i.e. a change of the derivative in the increasing behavior of Ta, induced by the persistent change of behavior of SSN which starts to affect the contribution of extrasolar components. The emissions of greenhouse gases in fact do not diminish and do not contribute to the behavior change of Ta. Finally, from 2007 to present (2015) the trend of Ta can be considered essentially stationary, without growth and therefore the effect of the solar component balance here the effects of the extra-solar components which tend to increase Ta.

\begin{figure}[!h]
\centering
\includegraphics[scale=1]{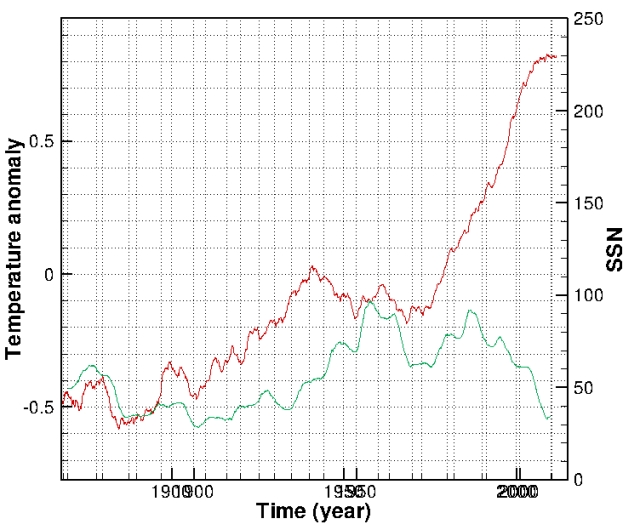}
 \caption{Red line: Ta series, filtered with a moving window of 96 months and 9 years shifted with respect to SSN series to account for the delay in the response of temperature.}
 \label{fig5}
\end{figure}

Moreover, we note that the current stationary trend of Ta is expected to continue for other several years (at least 10 years) on the basis of a prediction method obtained with an appropriate neural network model (NN), see Figure 6.

\begin{figure}[!h]
\centering
\includegraphics[scale=1]{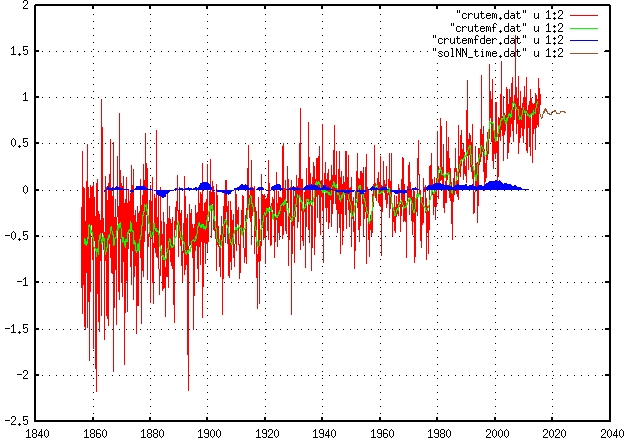}
 \caption{As in Figure 1, where the brown line corresponds to the 10 years prediction of the filtered temperature anomalies as obtained using a Neural Network prediction model.}
 \label{fig6}
\end{figure}

Thus, the above analysis shows that the contribution of the solar component on the behavior of global temperature anomalies is well confirmed and significant, enough to change the trend (i.e. the progressive growth at constant rate) of Ta after 14 years from the start of trend reversal of SSN and to stop its growth after 21 years.

A first approximation equation that linearly correlates the filtered SSN with the trend of the filtered Ta is given by: $Ta = 0.0066 \times SSN - 0.62 ( ^\circ C)$. This equation allows to single out the solar contribution approximated by the trend of temperature anomalies.

If this analysis will be confirmed, a further decrease of the SSN in the future might  be able to reverse the trend of temperature anomalies causing them to decline significantly for the first time since 1975 or, if the decrease of SSN should stop, the Ta should  remains in a stationary phase as the current one, at the same conditions on the anthropogenic or not contributions. If solar activity should instead increase in the future, among other factors, also the solar component should be added to the overall growth of Ta.
This analysis also shows that if in this period there was not present the mitigating action of the solar activity contribution, for different cycles in a decreasing trend, the actual evolution of temperature anomalies would be to increase continuously and quickly towards ever greater values, mainly due to greenhouse gases emission, due to human activities and natural processes. In the case of a restart increase of solar activity, it would be added also the solar contribution to the global temperature increase.

This result is in a substantial agreement with that recently found by Stauning, (2014), using different indicators of correlation between the time series of temperature anomalies and solar activity.

Finally, using the above formula of the approximate correlation between the series Ta and SSN in order to remove the contribution of the solar component on the temperature anomalies, we obtained the result shown in Figure 7.

\begin{figure}[!h]
\centering
\includegraphics[scale=1]{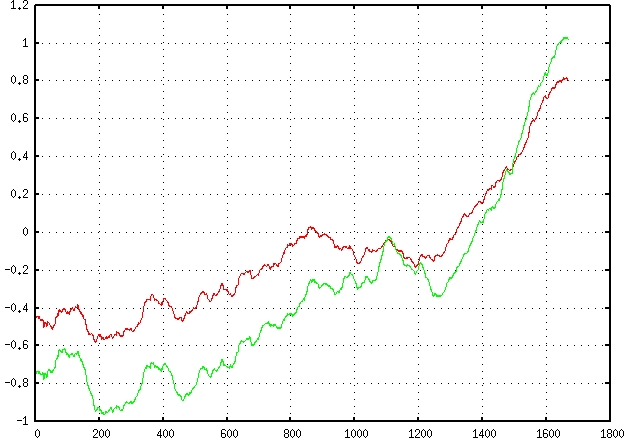}
 \caption{Red line: Global filtered temperature anomalies with solar contribution. Green line: Global filtered temperature anomalies without  solar contribution. Time is expressed in months after epoch 1870.}
 \label{fig7}
\end{figure}

As we can clearly see, with the lack of the solar contribution to mitigate the growth of temperature anomalies after 1986, it would have resulted in a greater and more rapid increase of Ta, to exceed the threshold of 1 degree of anomaly instead of the actual value of 0.8. Note that the two curves begin to separate and to diverge since 1993 well before the inflection point and about 8-9 years after the beginning of the reversal in the solar activity, confirming the time delay of the temperature anomalies response to solar forcing. Note also that these values should be considered only indicative, given the degree of approximation of the used formula for linear correlation.

\section{Conclusions}

By correlating the time series of global temperature anomalies and Solar Sunspot Number, properly filtered to determine the long-term trend, it is shown how the solar contribution affects the observed behavior of temperature anomalies. The solar component plays an important role and affects significantly the current stationary behavior of the global temperature anomalies. In particular, the analysis clearly shows that if there was absent in this period the mitigating action of reduced solar activity, for different last cycles in a decreasing trend, the actual evolution of temperature anomalies would be that of a continuous and quick increase towards increasing values, mainly due to emissions of greenhouse gases both by human activities and natural ones. In the case of an increasing solar activity, it would be added also the solar contribution to the rise of temperatures. It is therefore crucial the current objective of reducing the greenhouse gas emissions to avoid a future excessive increase in global temperatures, especially in the absence of the mitigating role of solar activity. In fact, its behavior and its future role will therefore be decisive in determining whether or not we will record a restart of the global temperature anomalies growth observed since 1975. \\ \\

\section{References}

Met Office Hadley Centre observations datasets, CRUTEM4 Diagnostics, \\  \ http://www.metoffice.gov.uk/hadobs/crutem4/data/diagnostics/\

SIDC-team, World Data Center for the Sunspot Index, Royal Observatory of Belgium, Monthly Report on the International Sunspot Number, online catalogue of the sunspot index: \ http://sidc.oma.be/silso/datafiles\

Feng-Ling Gao, Le-Ren Tao, Guo-Min Cui, Jia-Liang Xu, Tse-Chao Hua, 2015: The influence of solar spectral variations on global radiative balance, Advances in Space Research, Volume 55, Issue 2, 15 Pages 682 - 687

Lockwood M., 2012: Solar influence on global and regional climates, Surveys in Geophys., 33 3-4.

Scafetta, N., West, B.J., 2006: Phenomenological solar contribution to the 1900 - 2000 global surface warming. Geophys. Res. Lett. 33, L05708.

Scafetta, N., 2011: Total Solar Irradiance Satellite Composites and their Phenomenological Effect on Climate, in: Easterbrook, D. (Eds.), Evidence-Based Climate Science. Elsevier Inc., Amsterdam, pp. 289-316.

Sello, S., 2003: Wavelet entropy and the multi-peaked structure of solar cycle maximum, New Astronomy, Volume 8, Issue 2, p. 105-117

Stauning, P.: Reduced Solar Activity Disguises Global Temperature Rise, Atmospheric and Climate Sciences, 2014, 4, 60-63

WMO Report on Greenhouse Gas Concentrations, \\  \ http://www.wmo.int/pages/mediacentre/press\_releases/pr\_980\_en.html\

\end{document}